\documentclass[lettersize,journal]{IEEEtran}
\usepackage{amssymb}
\usepackage{caption}
\usepackage{multirow}
\usepackage{url}
\usepackage{booktabs}
\usepackage{pifont}
\usepackage{makecell}
\usepackage{verbatim}
\usepackage{breakcites}
\usepackage{forest}
\usepackage{tabularx}
\usepackage{booktabs}
\usepackage{caption}
\usepackage[english]{babel}
\usepackage[caption=false]{subfig}
\usepackage[noadjust]{cite}
\usepackage{cite}
\usepackage[numbers]{natbib}
\usepackage{algorithm}
\usepackage{algpseudocode}
\usepackage{booktabs,tabularx, array}
\renewcommand{\arraystretch}{1.2}

\usepackage{pifont}
\renewcommand{\arraystretch}{1.2}

\ifCLASSINFOpdf
\usepackage{graphicx}

\begin{document}
\title{Influence of Object Affordance on Action Language Understanding: Evidence from Dynamic Causal Modeling Analysis 
}

\author{Supriya Bordoloi$^{1}$, 
Cota Navin Gupta$^{1,2}$
\IEEEmembership{Member, IEEE},
and Shyamanta M. Hazarika$^{1,3}$
\IEEEmembership{Senior Member, IEEE},

\thanks{$^{1}$S. Bordoloi is affiliated with the Centre for Linguistic Science and Technology (CLST), Indian Institute of Technology Guwahati, 781039, India. email: 
        {bordoloi@iitg.ac.in}}
\thanks{$^{2}$C. N. Gupta is affiliated with the Neural Engineering Lab, Department of Bio Sciences and Bio Engineering, and with CLST, Indian Institute of Technology Guwahati, 781039, India. email: 
        {\small cngupta@iitg.ac.in}}%
\thanks{$^{3}$S. M. Hazarika is affiliated with Biomimetic Robotics and Artificial Intelligence Lab (BRAIL), Department of Mechanical Engineering, and with CLST, Indian Institute of Technology Guwahati, 781039, India. email: 
        {s.m.hazarika@iitg.ac.in}}%
        }

\maketitle
\begin{abstract}
This study investigates the causal neural dynamics by which affordance representations influence action language comprehension. In this study, 18 participants observed stimuli displayed in two conditions during the experiment: text-only (e.g., `Hit with a hammer') and video+text (visual clips with matching phrases). EEG data were recorded from 32 channels and analyzed for event-related potentials and source localization using LORETA, which identified four left-hemisphere regions of interest: the Lateral Occipital Cortex (LOC), Posterior Superior Temporal Gyrus (pSTG), Ventral Premotor Cortex (PMv), and Inferior Parietal Lobule (IPL). A space of dynamic causal modeling (DCM) was constructed with driving inputs to LOC and pSTG, and multiple connectivity configurations were tested. Bayesian Model Selection revealed a dominant model in which PMv causally influenced IPL and pSTG, reflecting a feedforward architecture from affordance-related motor regions to semantic hubs. Bayesian Model Averaging further confirmed strong endogenous connections from LOC to PMv and IPL, and significant modulation from PMv to IPL. These findings provide direct evidence that affordance processing in premotor regions drives action language understanding by engaging downstream parietal and temporal areas. The results support grounded cognition theories and offer a mechanistic account of how sensorimotor information contributes to linguistic comprehension.
\end{abstract}

\section{Introduction}
\label{sec:sample1}
\IEEEPARstart{U}{nderstanding} action-related language engages the brain regions responsible for perception and motor control. 
Action verbs that conveys the execution of a physical or mental action, offer insights into how language processing interacts with perception and motor experience.
Early cognitive models conceptualized language comprehension as an amodal and symbolic process, where meanings are represented and manipulated within an abstract, language-specific cognitive system that operates independently of sensory and motor modalities \cite{fodor1983modularity}. 
According to this view, understanding a sentence such as \textit{ `He grasped the handle'} does not necessitate the recruitment of perceptual or motor systems; instead, meaning is derived through rule-based manipulation of symbols that refer to representational forms in a disembodied fashion. These models often regard the brain's sensorimotor systems as peripheral to the core mechanisms of language understanding.
In contrast, embodied cognition theories challenge this disjunction by proposing that cognitive processes, including language comprehension, are fundamentally grounded in the body’s perceptual and motor engagement with the environment.
\cite{barsalou2008grounded, pulvermuller2005brain}. 
Within this framework, understanding action-related language involves the reactivation or simulation of sensorimotor experiences associated with the described actions. For instance, the verb `grasp', denoting an action, is understood not only through symbolic abstraction but also via the reactivation of motor experiences which is linked to actual grasping movements. \cite{pulvermuller2005brain, pulvermuller2005functional, zwaan2006}.

Neuroimaging and electrophysiological studies lend strong support to the embodied view, revealing that reading or listening to action-related verbs triggers motor and premotor cortical activation that follows a somatotopic organization, suggesting that the brain simulates the physical execution of actions during language comprehension \cite{hauk2004somatotopic, fischer2008embodied}. For instance, verbs describing leg, arm, or facial actions have been shown to activate corresponding regions in the motor cortex during passive reading tasks, indicating a direct mapping between linguistic input and sensorimotor representations \cite{pulvermuller2001walking, pulvermuller2001memory}. 
Importantly, recent research suggests that such activations may be modulated or even initiated by the brain's processing of object affordances, which are the action possibilities an object naturally offers. 
For example, the sight of a hammer can automatically evoke motor plans associated with hitting, which can prime or facilitate the comprehension of related action language \cite{grezes2003objects, gallivan2013decoding}. Behavioral and neural studies have shown that affordance processing can enhance the speed and accuracy of action verb comprehension, suggesting that the perception of functional object properties provides pre-linguistic, grounded cues that influence language understanding \cite{myung2006affordances, bub2018affordance}.

Behavioral studies have shown that objects with clear action possibilities can prime the interpretation of related action verbs, often speeding up reaction times and enhancing accuracy in language tasks \cite{tucker2004action, lindemann2006semantic}.  
Neurophysiological findings complement this by revealing that affordance-based cues can trigger preparatory motor responses even before explicit linguistic input is processed \cite{boulenger2006cross, mirabella2012object}. 
These patterns suggest that affordance perception may play a predictive or preparatory role in language understanding, providing sensorimotor context that facilitates subsequent semantic interpretation \cite{vanelk2014sensorsim, bub2018affordance}. 
The interaction between affordance perception and language understanding likely unfolds in sequence, helping the brain build meaning from both sensorimotor and linguistic input in a coordinated and grounded way \cite{glenberg2012action, myung2006affordances}. 
However, many earlier studies investigating neural responses to action language and affordance-based stimuli have largely relied on correlational measures, such as activation patterns or coherence between regions, which do not capture the directional influence between brain areas \cite{pulvermuller2005brain, hakan2006event}. For instance, although studies have examined superior temporal gyrus (STG) and motor cortex activity during audiovisual language tasks \cite{beauchamp2004integration, skipper2007hearing}, they often fail to explore the dynamic interactions underlying the integration of multimodal stimuli and the brain’s response to contextual affordances\cite{arnal2011multisensory, kaiser2013object}.

Dynamic Causal Modeling (DCM) addresses these limitations by offering a computational framework to identify causal interactions among brain regions from neurophysiological data, such as EEG, MEG, or fMRI. Unlike conventional connectivity analyses, DCM explicitly models how experimental conditions modulate directed (effective) connectivity between neural populations within a specified network \cite{friston2003dynamic, david2006dcm}. It does so by incorporating biophysical and neuronal models to estimate not just which areas are active, but how and when one area influences another in response to task-specific stimuli.
DCM was originally developed for fMRI but has since been adapted to EEG and MEG, enabling high-temporal-resolution analysis of rapid neural dynamics. This is particularly valuable for paradigms involving rapid stimulus presentations, 
where changes in connectivity may unfold over hundreds of milliseconds \cite{kiebel2009dynamic, stephan2010ten}. For instance, in speech perception, DCM has been used to reveal top-down modulatory effects from frontal to temporal areas during comprehension of degraded or ambiguous stimuli \cite{garrido2007evoked, gao2020dynamic}.
In recent years, DCM has also been applied to action language paradigms. For instance, Maess et al. \cite{maess2020language} showed that motor and auditory regions form an interactive network during action word processing using MEG signals. 
Similarly, Straube et al. \cite{straube2013supramodal} demonstrated supramodal integration between motor and language systems using fMRI during gesture-speech integration . Other studies have extended DCM to explore how contextual, semantic, and embodied variables modulate effective connectivity in language tasks, revealing flexible recruitment of motor areas depending on task demands \cite{willems2010interaction, molle2016cortico}. Findings by Chang et al. \cite{chang2018modulation} and Ghosh et al. \cite{ghosh2012inverse}  indicated that understanding action verbs involves the dynamic recruitment of sensorimotor networks, particularly when stimuli are ambiguous or involve higher semantic load. Despite these advances, most of the existing work has concentrated on symbolic language or gesture processing and does not directly investigate how object affordances, understood as the action possibilities suggested by objects, may causally influence the semantic interpretation of action verbs.
Despite these advances, most prior research has treated affordance and language as parallel or simultaneously activated processes, without explicitly modeling the direction or timing of information flow between them. As a result, the temporal dynamics and causal structure underlying how affordance cues influence language comprehension remain insufficiently understood. 
There is a critical need to explore how sensorimotor systems contribute to semantic processing. To address this, we examined whether affordance-related regions actively drive semantic networks during action language comprehension, thereby supporting a grounded, sensorimotor account of language understanding. 

This study employed an event-related EEG experiment to investigate the causal influence of object affordance on action language processing. EEG recordings were obtained to investigate the dynamic interplay between motor and parietal cortices during affordance-based language comprehension. We hypothesized that object affordance representations activated during perception causally influence brain regions involved in understanding action language. 
To test our hypothesis, we defined four key regions of interest (ROIs): the lateral occipital cortex (LOC), posterior superior temporal gyrus (pSTG), ventral premotor cortex (PMv),  and inferior parietal lobule (IPL). Based on these ROIs, we constructed a model space consisting of 35 candidate models that reflect different plausible configurations of effective connectivity between these regions. We employed DCM for ERP data to estimate and compare the models. Bayesian Model Selection (BMS) was performed at the group level using a random-effects approach to identify the most likely model, followed by Bayesian Model Averaging (BMA) to infer the connectivity architecture across the model space. This framework enables a mechanistic understanding of how object affordance modulates the neural network underlying action language comprehension, and supports causal inferences beyond correlational analysis.

\section{Materials and Methods}
\subsection{Demographic Information}

\begin{figure*}[h!]
 	\footnotesize
 	\centering
 	\includegraphics[scale=0.30]{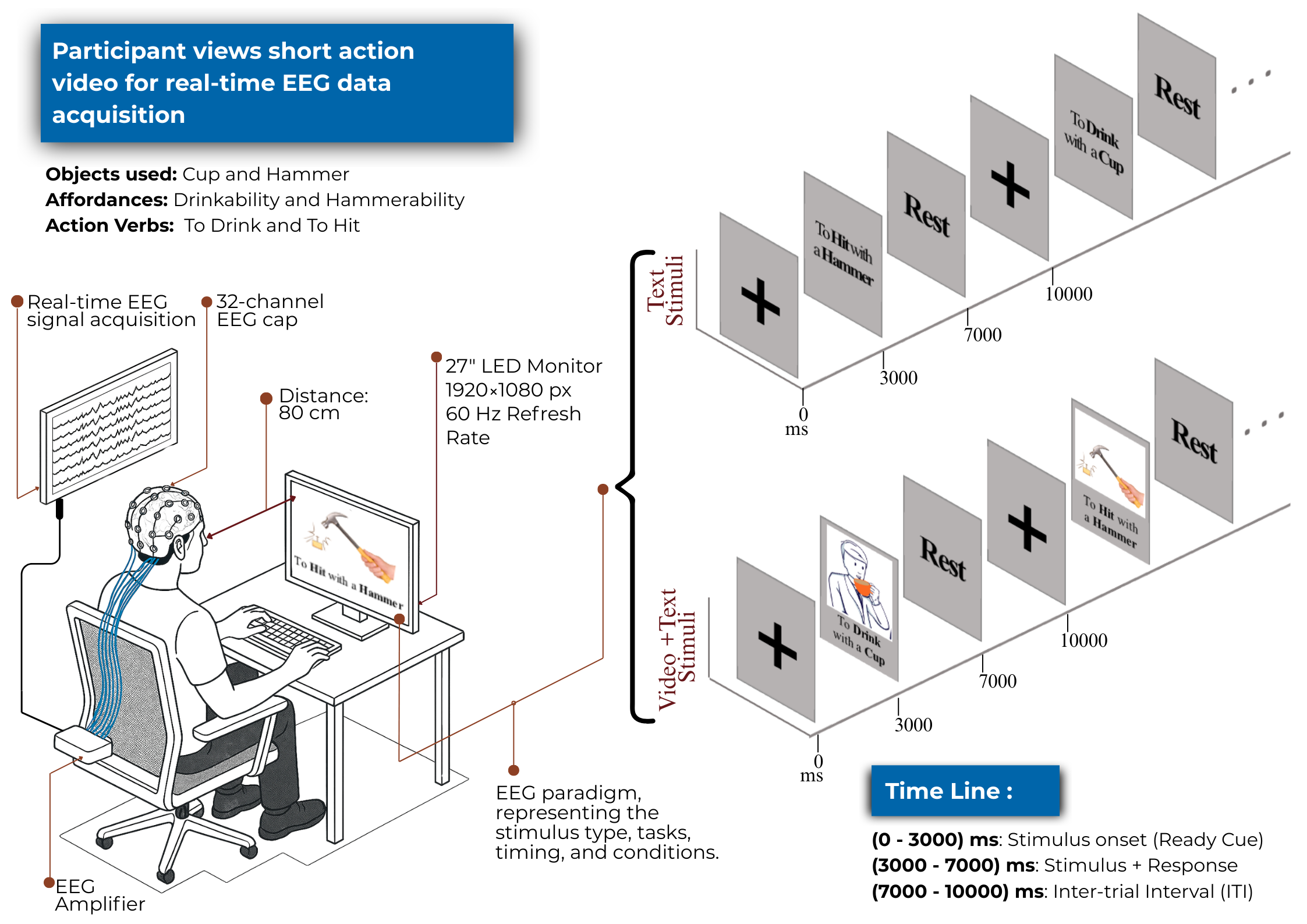}
  \caption{Experimental paradigm with two conditions: a. Text-only stimuli, b. Video+Text stimuli. Each condition included 20 trials, with each trial lasting 10000 ms. A fixation cross was presented during the initial 3000 ms of each trial. In the first Condition, participants were shown a textual description of an affordance-driven action verb for 4000 ms. The next 3000 ms was a resting period or inter-trial interval (ITI) where participants were allowed eye blinking and relaxation. In the second condition, subsequent to the presentation of a fixation cross, a video depicting the same action is displayed along with the corresponding text. This again lasts for 4000 ms. An ITI of 3000 ms follows.} 
 	\label{paradigm}
 \end{figure*}

18 right-handed and healthy students from IIT Guwahati (age $31.0$ $\pm$ $2.66$), participated voluntarily in the current EEG experiment. None of the participants reported any psychiatric or neurological disorders, and all had normal or corrected-to-normal vision. The study was approved by the Ethics Committee of IIT Guwahati vide reference no. IHEC/SH/11/2023. Prior to participation, every subject gave informed written consent.

\subsection{Expermental Design}

Semantically congruent stimuli were administered to participants in two conditions: 
(1) text-only and (2) video+text. Each condition included 20 trials, with each trial lasting 10000 ms. A fixation cross was presented during the initial 3000 ms of each trial, as shown in Fig. \ref{paradigm}. 
In the text-only condition, stimuli comprised short action sentences in infinitive form that explicitly described object affordances, such as `To Hit with a Hammer' and `To Drink with a Cup'. These sentences were presented visually in isolation. In the video+text condition, participants viewed short animated video clips depicting meaningful object–action pairings (e.g., a person drinking from a cup or hitting with a hammer), simultaneously accompanied by the corresponding sentences. 
All experimental trials included stimuli that were presented in a randomized and counterbalanced order.
\subsection{EEG Measurement and Analysis}
A 32-channel EEG setup with Ag/AgCl electrodes was employed, positioned in line with the international 10–20 system, covering the following locations.: Fp1, Fp2, F3, F4, Fz, FC1, FC2, FC5, FC6, F7, F8, FT9, FT10, CP1, CP2, C3, C4, Cz, CP5, CP6, P3, P4, Pz, P7, P8, O1, O2, T7, T8, TP9, TP10, and VEOG. 
EEG signals were collected with the BrainVision Recorder system \cite{brainproducts2025-misc} at a 500 Hz sampling frequency. Impedance levels of all electrodes were controlled to remain under 10 k$\Omega$ during recording.
Pre-processing was performed using BrainVision Analyzer. The dataset was downsampled to 256 Hz and filtered with a 0.1–40 Hz bandpass. Power line artifacts at 50 Hz were eliminated using a notch filter, while a zero-phase Butterworth filter ensured that the temporal structure of the signals was retained.
An offline average reference, excluding the VEOG channel, was applied to the EEG data. Subsequently, continuous signals were divided into epochs spanning –200 ms to 0 ms relative to stimulus onset, covering both text-only and video+text trials.
Signals were corrected to baseline by referencing the –200 to 0 ms window before stimulus onset. Artifact rejection and ocular correction procedures were implemented using the BrainVision Analyzer to ensure high data quality prior to further analysis.
\begin{figure*}[h!]
 	\footnotesize
 	\centering 	\includegraphics[scale=0.30]{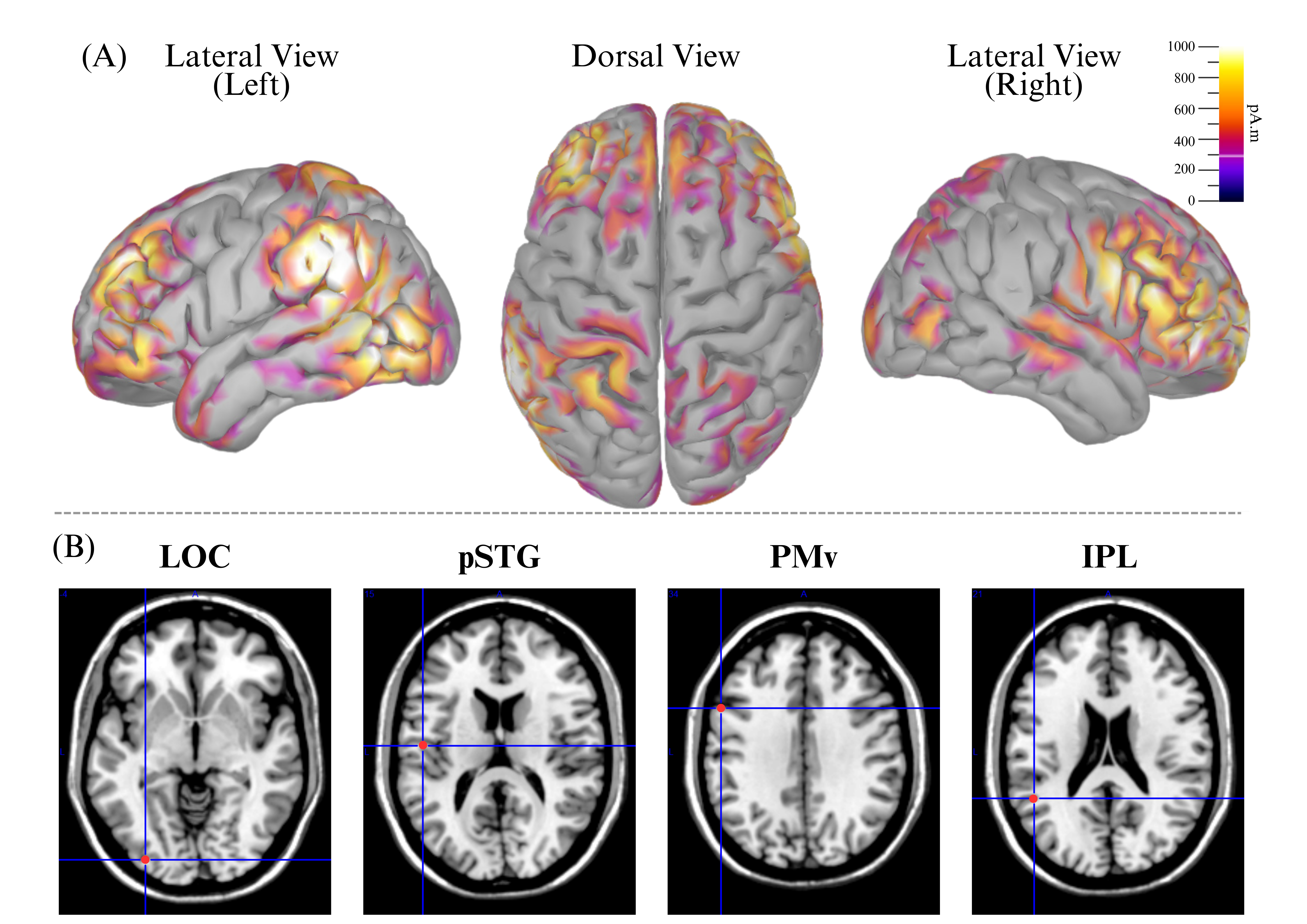}
  \caption{Condition-specific brain activation: (A) Whole-brain surface renders of the group t-map, showing task-related activity distributed over occipital, temporal, parietal, and premotor regions. (B) Axial slices centered on the ROIs: LOC, IPL, pSTG, and PMv, highlighting the peak voxels. Images are shown in MNI space.
  } 
 	\label{roi}
 \end{figure*}  
The preprocessed EEG epochs were analyzed to extract Event-related potential (ERP)  elements corresponding to the two conditions. 
Based on the cognitive processes involved in semantic integration and object affordance perception, two primary ERP components were analyzed: N200 and N400.
The N200 component (150–200 ms post-stimulus) is associated with early attention allocation, affordance processing, and processing of visual and language features, especially when related to meaning or object use \cite{rowe2017objects}. This component was analyzed primarily over fronto-central electrode sites to capture early affordance-related differentiation in both the text-only and video+text conditions. 
The N400 component (300–500 ms post-stimulus) was selected as the primary index of semantic integration and action-language comprehension. It is well-established that N400 amplitude reflects the incorporation of word meaning into context and is influenced by action-related vocabulary \cite{kutas2011thirty, proverbio2009rp, debruille2007n400}.  
For both conditions, N400 activity was examined at centro-parietal regions, as these have been consistently associated with action-semantic processing \cite{sitnikova2008two}. The identification of these ERP components and their respective time windows was guided by grand-average ERP waveforms across participants.

\subsection{Source Analysis} 

To localize neural generators involved in affordance-driven action language processing, source-level analysis was performed in two stages. First, we used Brainstorm software \cite{Tadel2011} to visualize overall cortical activation patterns across all subjects and conditions [see Fig. \ref{roi}(A)]. Grand-average source maps were computed. 
Next, to precisely identify the anatomical locations of key sources, we took advantage of the Low-Resolution Brain Electromagnetic Tomography (LORETA) module in BrainVision Analyzer. Source estimation was performed separately for the text-only and video+text conditions within two ERP-defined time windows: 150–250 ms (N200) and 300–500 ms (N400). This analysis revealed peak activations in four left-hemisphere regions: LOC, pSTG, PMv, and IPL. These four regions were selected as regions of interest (ROIs) for a subsequent DCM investigation. The localized ROIs are illustrated in Fig. \ref{roi}(B).

\section{Dynamic Causal Modeling Analysis} 
We employed DCM to examine the directional connectivity associated with affordance-driven action language processing, using the SPM12 toolbox \cite{SPM12}. DCM estimates directed (causal) interactions between neuronal sources by fitting a biophysically informed generative model to the observed EEG data \cite{friston2003dynamic, Kiebel2006}. 
\subsubsection{Regions of interest (ROI) selection }
\begin{table}[t]
\centering
\caption{Peak MNI coordinates and $t$-values for ROIs displayed in Fig.~2.}
\label{tab:roi_coords}
\begin{tabularx}{\columnwidth}{
  >{\raggedright\arraybackslash}X 
  *{3}{>{\centering\arraybackslash}X}
  >{\centering\arraybackslash}X       
}
\toprule
\textbf{Region of Interest} & \multicolumn{3}{c}{\textbf{MNI coordinates}} & \textbf{$t$-value} \\
\cmidrule(lr){2-4}
& \textbf{x} & \textbf{y} & \textbf{z} & \\
\midrule
LOC  & $-40$ & $-66$ & $-8$  & 4.9 \\
IPL  & $-48$ & $-64$ & 30    & 6.1 \\
pSTG & $-66$ & $-38$ & 22    & 5.3 \\
PMv  & $-58$ & 13    & 19    & 5.0 \\
\bottomrule
\end{tabularx}
\end{table}

To define a biologically plausible model space for Dynamic Causal Modeling, four cortical ROIs were chosen based on a combination of extensive literature review and empirical source-level findings obtained via LORETA. MNI co-ordinates of these ROIs are  shown in Table \ref{tab:roi_coords}. These ROIs were chosen to represent key nodes involved in visual affordance processing, sensorimotor integration, and action language comprehension, which are the core components of our experimental hypothesis. 
As the primary visual region for object processing, LOC was chosen because of its involvement in extracting object shape and manipulability features. These outputs are foundational for downstream sensorimotor and semantic processing \cite{Palejwala2020LOC, Chen2022ManipulabilityEbbinghaus}. 
pSTG is fundamental to motor task performance by serving as a multisensory integrative hub for the processing of object affordances and the comprehension of action language \cite{zschorlich2021multimodal, buccino2004neural}. 
PMv plays a crucial role in motor tasks by representing object affordances through visuomotor transformations and integrating object-related visual information \cite{Maranesi2014}. It also contributes causally to the processing and encoding of action language from textual stimuli, especially for manual actions \cite{Tremblay2011}. 
Lastly, IPL was incorporated due to its selective response to manipulable objects and its critical involvement in linking visual inputs to potential motor actions. IPL also serves as an interface between affordance perception and action-related language networks \cite{Chao2000, zwaan2006seeing}. 

\begin{figure*}[h!]
 	\footnotesize
 	\centering 	\includegraphics[scale=0.29]{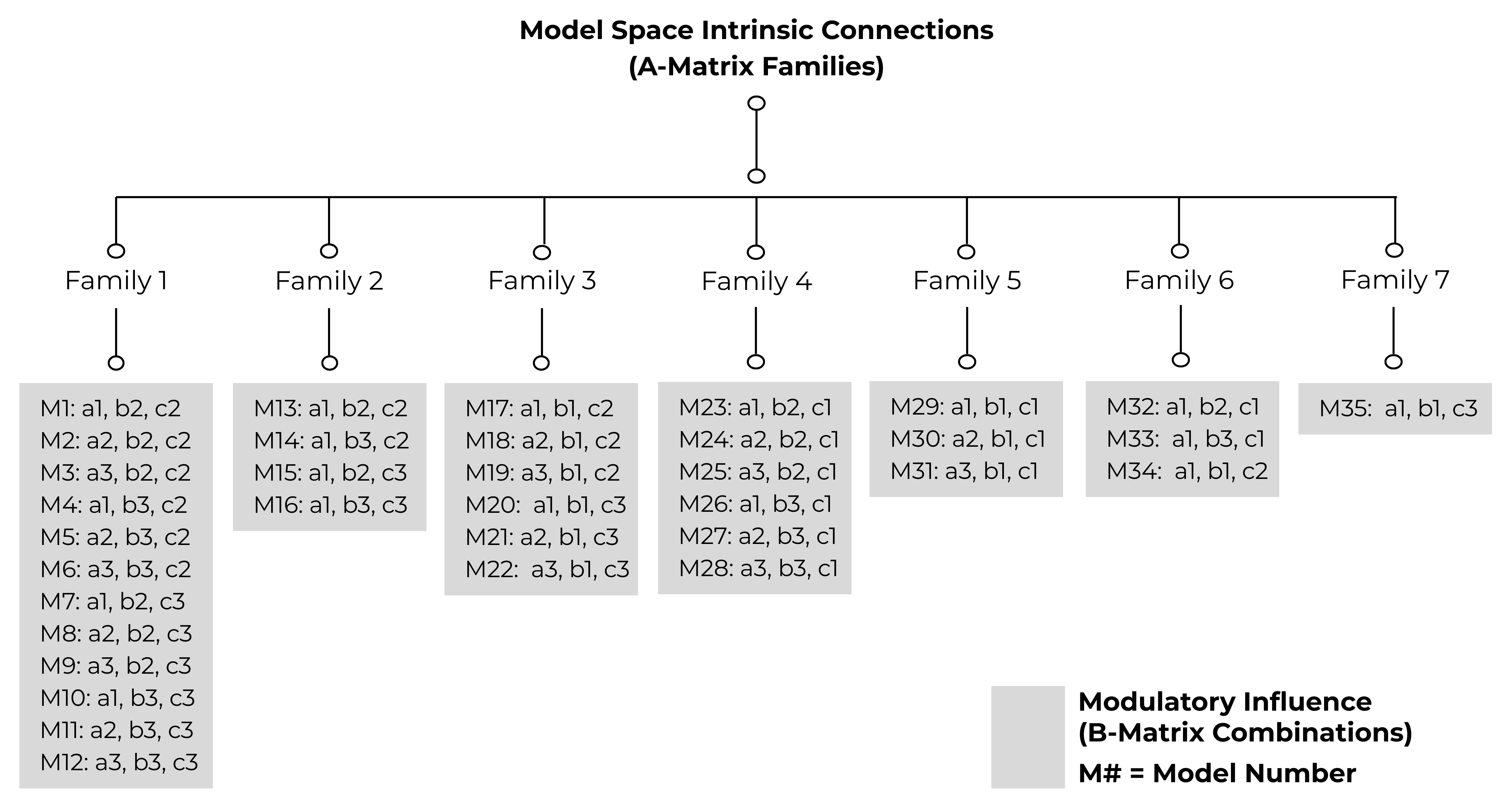} 
  \caption{Model space consisting of Intrinsic connections (A-matrix families) and modulatory influences (B-matrix combinations) 
  } 
 	\label{model}
 \end{figure*} 
 \begin{figure*}[h!]
 	\footnotesize
 	\centering 	\includegraphics[scale=0.28]{Fig_04.png}
  \caption{DCM model space and families. The schematic shows four ROIs—PMv, IPL, pSTG, and LOC—with fixed endogenous connections (black) and constant driving inputs to LOC and pSTG (dotted). Red arrows mark family-specific modulatory pathways. Bottom panels show one representative model from each of the seven families (35 models total) used for Bayesian model comparison.} 
 	\label{models_fig4}
 \end{figure*} 
\subsection{Model Space Construction}

Based on the four selected ROIs: LOC, pSTG, PMv, AND IPL, we constructed a set of DCMs to test our hypothesis regarding the directionality and modulation of connectivity underlying affordance-driven action language processing. The model space was structured to explore variations in both endogenous (intrinsic) connections and context-dependent modulatory influences, while keeping the driving inputs (C-matrix) fixed across all models (Eq~\ref{eq:dcm}).
\begin{equation}
\dot{w}(t) = \left( A + \sum_{j=1}^{m} v_j(t) B^{(j)} \right) w(t) + C v(t)
\label{eq:dcm}
\end{equation}
Here,
\begin{itemize}
    \item \( w(t) \) is the neural state vector representing activity in each ROI at time \( t \).
    \item \( A \) is the matrix of intrinsic (fixed) or context-invariant links across the ROIs.
    \item \( B^{(j)} \) represents modulatory (condition-specific) effects on connectivity induced by experimental condition \( j \).
    \item \( v_j(t) \) represents a binary input variable that specifies whether the condition \( j \) is present or absent at time \( t \).
    \item \( C \) defines the driving inputs (e.g., stimuli) that directly influence specific ROIs.
    \item \( v(t) \) is the external input vector (e.g., text-only or video+text stimuli).
\end{itemize}
To systematically generate plausible model configurations, we implemented a model space generation routine, as shown in Fig \ref{model}. It takes a list of user-defined regions of interest (ROIs) and fixed endogenous connections to construct a connectivity matrix. It then identifies all non-endogenous pairs and exhaustively generates combinations of modulation possibilities, resulting in a comprehensive list of DCM model families.

\subsubsection{Intrinsic Connections (A-matrix)}
The intrinsic connections focused on the pattern of endogenous connectivity among the four ROIs. Bidirectional connections were specified between pSTG$\leftrightarrow$LOC, as evidenced by studies demonstrating that language-related symptoms are associated with connectivity linking the superior temporal gyrus and occipital cortex (of which LOC is a part) \cite{gaebler2023functional}. Unidirectional connections from LOC$\rightarrow$IPL and LOC$\rightarrow$PMv were also included, consistent with the dorsal stream framework, in which object-related visual information flows from occipital to parietal and frontal regions for affordance extraction and motor simulation \cite{borghi2015stable, wurm2017action}. Additionally, connections involving PMv$\leftrightarrow$pSTG, PMv$\leftrightarrow$IPL, and IPL$\leftrightarrow$pSTG were varied across models to explore how premotor and parietal areas contribute to semantic and motor aspects of affordance-language integration \cite{zwaan2006seeing, pulvermuller2013neurons, garcea2014parcellation}.
Following Friston et al. \cite{friston2003dynamic}, the basic architecture of these intrinsic connections was held constant in each family, but the configuration of specific connections varied across seven model families to test our hypothesis.

\begin{itemize}
  \item Family 1: PMv$\leftrightarrow$pSTG, PMv$\leftrightarrow$IPL, IPL$\leftrightarrow$pSTG
  \item Family 2: PMv$\leftrightarrow$pSTG, PMv$\leftrightarrow$IPL
  \item Family 3: PMv$\leftrightarrow$IPL, IPL$\leftrightarrow$pSTG
  \item Family 4: PMv$\leftrightarrow$pSTG, IPL$\leftrightarrow$pSTG
  \item Family 5: IPL$\leftrightarrow$pSTG only
  \item Family 6: PMv$\leftrightarrow$pSTG only
  \item Family 7: PMv$\leftrightarrow$IPL only
\end{itemize}

\subsubsection{Modulatory Influences (B-matrix)}

Modulatory influences were modeled on selected connections to examine potential condition-specific effects during affordance-language integration. Specifically, we tested the following connection pairs for possible unidirectional, bidirectional, or absent modulation. 
For the pSTG$\leftrightarrow$IPL connection, three possibilities were modeled: no modulation, unidirectional modulation from pSTG$\rightarrow$IPL, and bidirectional modulation (pSTG$\leftrightarrow$IPL). Similarly, the connection between pSTG and PMv included models with no modulation, unidirectional modulation from pSTG$\rightarrow$PMv, and bidirectional modulation (pSTG$\leftrightarrow$PMv). Lastly, for the PMv$\leftrightarrow$IPL pathway, we tested no modulation, unidirectional modulation from PMv$\rightarrow$IPL, and bidirectional modulation (PMv$\leftrightarrow$IPL). 
Graphical representation of all plausible models is included in Fig. S1 as supplementary material.

\subsubsection{Driving Inputs (C-matrix)} 
All models assumed fixed driving inputs to pSTG and LOC, representing the points of entry for semantic and visual information, respectively. These inputs were consistent across all model families and allowed us to examine how the system responds to affordance-driven stimuli involving both text and visual cues. A schematic diagram of all families and their model spaces is shown in Fig. \ref{models_fig4}, with intrinsic connections, driving inputs, and modulatory pathways.

\subsection{Bayesian Model Comparison (BMC)}
To determine the most plausible connectivity architecture underlying affordance-driven action language processing, we conducted Bayesian Model Selection (BMS). This approach evaluates competing models by estimating the model evidence for each and comparing their relative fit to the observed ERP data across subjects.
DCMs were estimated for each participant individually using the variational Bayesian inversion scheme implemented in SPM12. The ERP data used for model inversion were time-locked to stimulus onset.
Following model inversion, random-effects BMS was applied across participants to account for inter-subject variability in the underlying neural architectures. The exceedance probability (EP) and expected posterior probability (EPP) were computed for each model and model family. The model (or family) with the highest EP was regarded as providing the most plausible explanation of the observed data at the group level. 
To assess which aspects of the model architecture were consistently supported, we also employed Bayesian model averaging (BMA). 
This allowed us to estimate average effective connectivity parameters across models, weighted by their posterior probabilities, and to identify connections that contributed most reliably to affordance-driven processing. 
This BMC framework enables us to identify the most likely causal architecture underlying how affordance information, processed through visual and sensorimotor regions, drives semantic comprehension in the brain.

\begin{figure*}[h!]
\centering
\includegraphics[scale=0.3]{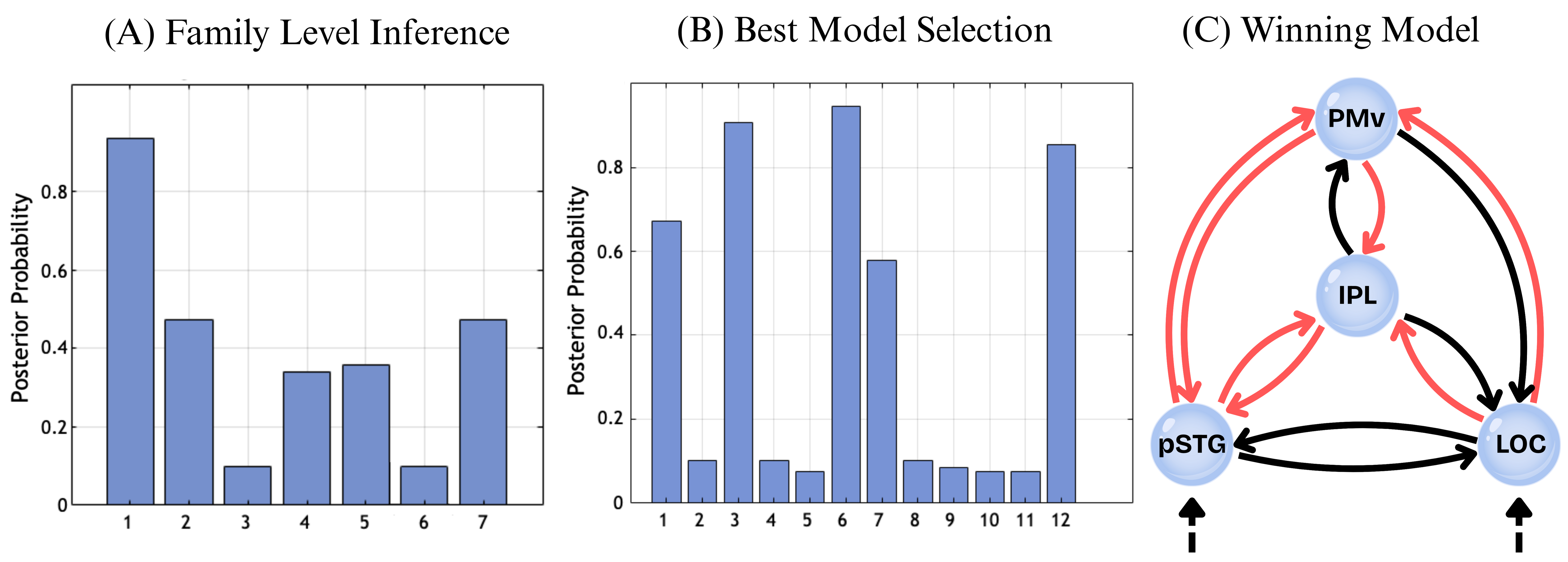}
\caption{(A) Family-level BMS showing Family 1 as the winning model group. (B) Model-level selection within Family 1, with M6 showing the highest log-evidence. (C) Winning model schematic showing key B-matrix modulations (PMv$\rightarrow$IPL, LOC$\rightarrow$IPL, pSTG$\rightarrow$PMv)}
\label{fig:bms}
\end{figure*}

\begin{figure*}[h!]
\centering
\includegraphics[scale=0.29]{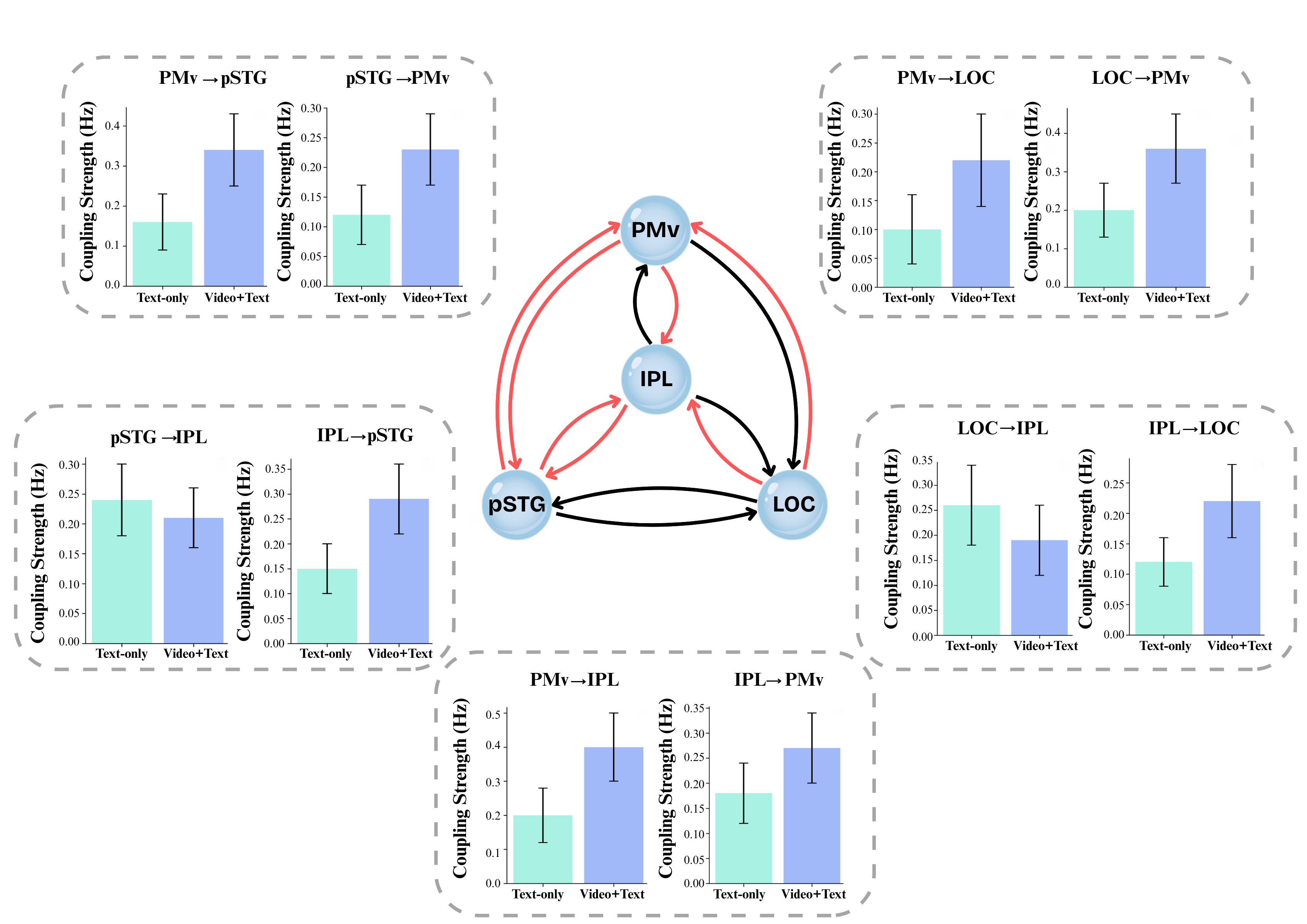}
\caption{Graphical representation of the winning DCM model (M6) showing the effective connectivity between four key ROIs: LOC, pSTG, PMv, and IPL. Bar chart showing estimated coupling strengths for each directed connection under the text-only and video+text conditions. The coupling values represent group-level Bayesian Model Averaging (BMA) estimates derived from the winning model, highlighting the stronger PMv$\rightarrow$IPL and PMv$\rightarrow$pSTG connections in the video+text condition compared to the text-only condition.}
\label{fig:coupling}
\end{figure*}

\section{Results}

\subsection{DCM model selection} 
Random-effects Bayesian Model Selection (RFX-BMS) at the group level favored the fully connected family (bidirectional intrinsic connections among LOC, pSTG, PMv, and IPL), with a posterior probability and EP (family) of $0.87$. Within this family, Model~6 (M6) showed the highest support, with a posterior probability and EP of $0.91$, indicating it was the most likely model to explain the observed data (Fig. \ref{fig:bms}). M6 specified modulatory influences during the video+text condition from PMv$\rightarrow$\{pSTG, IPL\}, IPL$\rightarrow$\{pSTG\}, LOC$\rightarrow$\{PMv, IPL\}, and pSTG$\rightarrow$\{PMv, IPL\}. The C-matrix specified driving inputs to LOC and pSTG.

BMA was performed across Family 1 to account for model uncertainty. The B-matrix results in Table~\ref{tab:Bmatrix} show that PMv$\rightarrow$IPL exhibited the strongest condition-specific modulation under video+text condition (mean = 0.28 Hz, SD = 0.05).
\begin{table}[htbp]
\centering
\caption{Modulatory Effects (B-matrix) (M6)}
\label{tab:Bmatrix}
\begin{tabular}{|c|c|c|c|c|}
\hline
\textbf{From / To} & \textbf{LOC } & \textbf{pSTG } & \textbf{PMv } & \textbf{IPL } \\
\hline
\textbf{LOC }   & --             & 0.16 $\pm$ 0.04 & --             & -- \\
\textbf{pSTG }  & 0.15 $\pm$ 0.03 & --              & 0.13 $\pm$ 0.04 & 0.10 $\pm$ 0.03 \\
\textbf{PMv }   & 0.14 $\pm$ 0.05 & 0.11 $\pm$ 0.04 & --             & 0.28 $\pm$ 0.05 \\
\textbf{IPL}   & 0.09 $\pm$ 0.03 & 0.12 $\pm$ 0.04 & --             & -- \\
\hline
\end{tabular}
\end{table}
The intrinsic (baseline) connectivity structure was revealed through the A-matrix (Table~\ref{tab:Amatrix}). PMv$\rightarrow$IPL and LOC$\rightarrow$IPL showed strong endogenous coupling, suggesting a baseline fronto-parietal pathway supporting affordance-driven comprehension.
\begin{table}[htbp]
\centering
\caption{Intrinsic Connectivity (A-matrix) between both Conditions}
\label{tab:Amatrix}
\begin{tabular}{|c|c|c|c|c|}
\hline
\textbf{From / To} & \textbf{LOC } & \textbf{pSTG } & \textbf{PMv } & \textbf{IPL } \\
\hline
\textbf{LOC }   & --    & 0.06 $\pm$ 0.01 & 0.02 $\pm$ 0.01 & 0.15 $\pm$ 0.03 \\
\textbf{pSTG }  & 0.03 $\pm$ 0.01 & --  & 0.11 $\pm$ 0.02 & 0.09 $\pm$ 0.02 \\
\textbf{PMv }   & 0.01 $\pm$ 0.01 & 0.05 $\pm$ 0.01 & --    & 0.21 $\pm$ 0.03 \\
\textbf{IPL }   & -0.01 $\pm$ 0.01 & 0.02 $\pm$ 0.01 & 0.00 $\pm$ 0.00 & -- \\
\hline
\end{tabular}
\end{table}
A significant negative correlation was observed between the subject-wise modulation strength of PMv$\rightarrow$IPL and response times ($r = -0.64$, $p = 0.01$). This indicates that greater fronto-parietal modulation facilitated faster semantic responses during video+text trials.

\subsection{DCM parameter statistics (RFX-BMA)}
Subject-specific parameters were obtained using random-effects Bayesian Model Averaging (RFX-BMA) across models in the winning family. 
Couplings are expressed as posterior means in Hz, estimated using RFX-BMA over models within the winning family. Classical one-sample $t$-tests (two-tailed) assessed whether couplings differed from zero; $P$-values were adjusted using the Benjamini–Hochberg FDR. Table~\ref{tab:DCMparams} summarizes intrinsic (A-matrix), modulatory (B-matrix), and resulting coupling (A$+$B). Values are posterior means (Hz)~$\pm$~SD.

\renewcommand{\arraystretch}{1.5}
\begin{table*}[t]
\centering
\caption{DCM parameters for the winning EEG model. Intrinsic (A), Modulatory (B; Video+text condition relative to text-only condition), and Resulting coupling (A$+$B). $t$-tests are one-sample against~0 (df = 17). FDR-corrected $P$ shown. Correlations (Spearman $\rho$) relate modulatory strength to behavioral indices: Accuracy gain 
and RT gain. ($P_{\mathrm{FDR}}$ values are Benjamini–Hochberg corrected. Values reported as $< .001$ reflect significance at $p < .001$ after correction.)}
\label{tab:DCMparams}
\resizebox{\textwidth}{!}{%
\begin{tabular}{lccccccccc}
\toprule
\multirow{2}{*}{\textbf{Connection (from$\rightarrow$to)}} & \multicolumn{3}{c}{\textbf{Intrinsic A (Hz)}} & \multicolumn{3}{c}{\textbf{Modulatory B }} & \multicolumn{3}{c}{\textbf{Resulting A$+$B} } \\
\cmidrule(lr){2-4}\cmidrule(lr){5-7}\cmidrule(lr){8-10}
& \textbf{Posterior Mean$\pm$SD} & \textbf{$t$} & \textbf{$P_{\mathrm{FDR}}$} & \textbf{Posterior Mean$\pm$SD} & $t$ & $P_{\mathrm{FDR}}$ & \textbf{Posterior Mean$\pm$SD} & $t$ & $P_{\mathrm{FDR}}$ \\
\midrule
PMv$\rightarrow$IPL       & $0.20\pm0.10$ & 8.9  & $<.001$ & \textbf{$0.20\pm0.09$} & 9.7 & \textbf{$<.001$} & $0.40\pm0.13$ & 13.5 & $<.001$ \\
PMv$\rightarrow$pSTG      & $0.16\pm0.09$ & 8.0  & $<.001$ & \textbf{$0.18\pm0.10$} & 8.5 & \textbf{$<.001$} & $0.34\pm0.12$ & 12.3 & $<.001$ \\
PMv$\rightarrow$LOC       & $0.10\pm0.09$ & 4.9  & $<.001$ & \textbf{$0.12\pm0.08$} & 6.8 & \textbf{$<.001$} & $0.22\pm0.11$ & 9.1  & $<.001$ \\
IPL$\rightarrow$pSTG      & $0.15\pm0.08$ & 8.4  & $<.001$ & \textbf{$0.14\pm0.08$} & 7.9 & \textbf{$<.001$} & $0.29\pm0.11$ & 11.3 & $<.001$ \\
IPL$\rightarrow$LOC       & $0.12\pm0.08$ & 6.7  & $<.001$ & \textbf{$0.10\pm0.07$} & 6.1 & \textbf{$<.001$} & $0.22\pm0.10$ & 9.6  & $<.001$ \\
LOC$\rightarrow$pSTG      & $0.20\pm0.10$ & 8.9  & $<.001$ & \textbf{$0.16\pm0.08$} & 8.7 & \textbf{$<.001$} & $0.36\pm0.12$ & 12.9 & $<.001$ \\
pSTG$\rightarrow$PMv      & $0.12\pm0.07$ & 7.5  & $<.001$ & \textbf{$0.11\pm0.07$} & 7.2 & \textbf{$<.001$} & $0.23\pm0.10$ & 10.3 & $<.001$ \\
pSTG$\rightarrow$LOC      & $0.18\pm0.09$ & 9.0  & $<.001$ & \textbf{$0.09\pm0.06$} & 6.6 & \textbf{$<.001$} & $0.27\pm0.11$ & 11.0 & $<.001$ \\
pSTG$\rightarrow$IPL      & $0.14\pm0.08$ & 7.6  & $<.001$ & \textbf{$0.08\pm0.06$} & 5.9 & \textbf{$<.001$} & $0.22\pm0.10$ & 9.8  & $<.001$ \\

\bottomrule
\end{tabular}
}
\vspace{0.25em}
\end{table*}

\begin{figure*}[h!]
\centering
\includegraphics[scale=0.34]{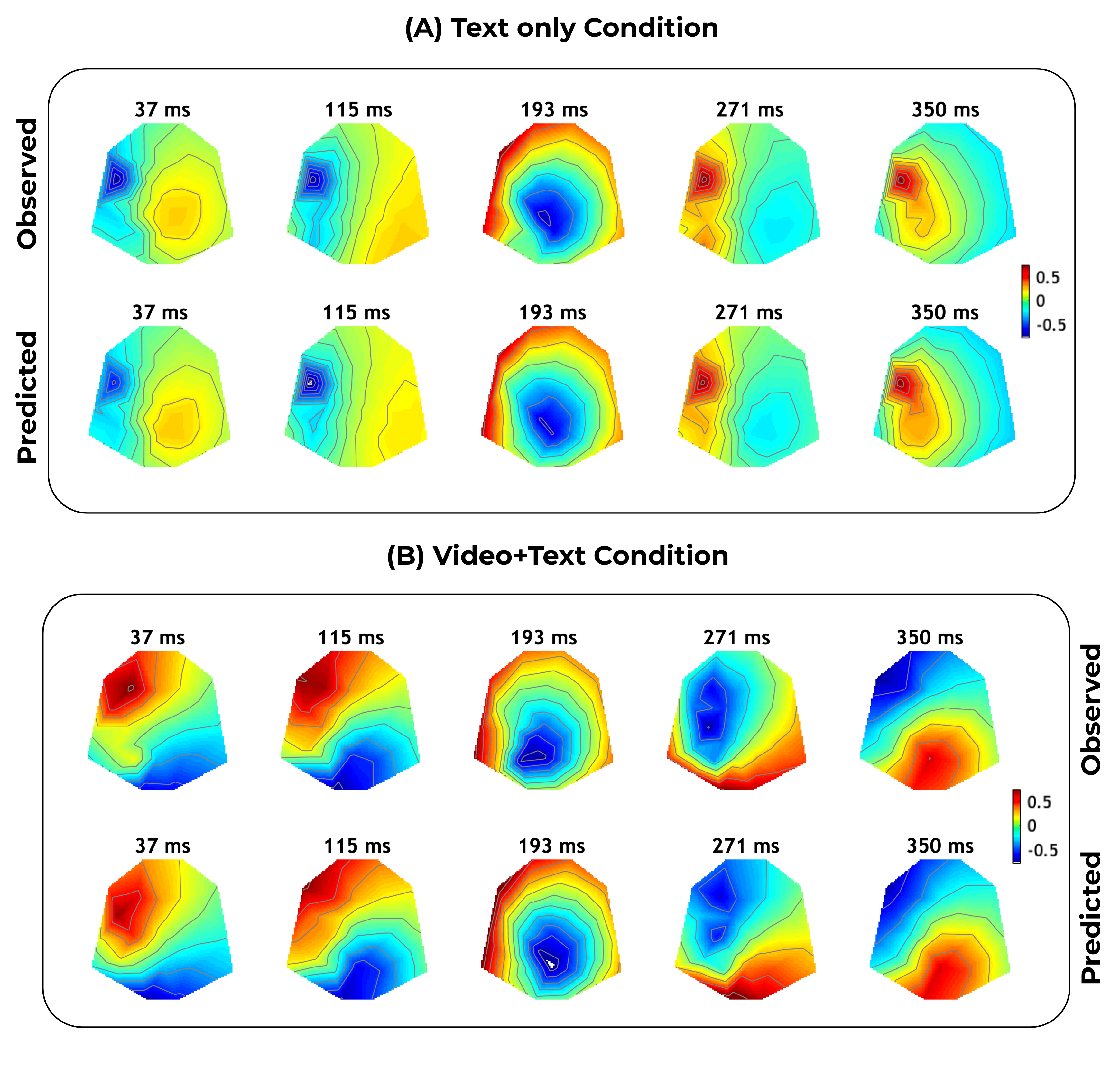}
\caption{Comparison of observed and predicted scalp topographies for text-only and video+text conditions. The figure illustrates the spatio-temporal distribution of EEG activity at representative 0-400 ms for both experimental conditions. The observed topographies are computed directly from the empirical EEG data, while the predicted maps are derived from the winning DCM model.}
\label{fig:topo}
\end{figure*}
\subsection{Connection-by-connection interpretation} 
The video+text condition selectively increased gain along PMv$\rightarrow$IPL, PMv$\rightarrow$pSTG, and PMv$\rightarrow$LOC, consistent with top-down motor predictions enhancing parietal affordance and sensory pathways. Feedforward LOC$\rightarrow$pSTG and IPL$\rightarrow$pSTG increased during video+text condition, indicating greater visual/parietal contributions to posterior temporal integration. Reciprocal pSTG$\rightarrow$\{PMv, LOC, IPL\} modulations suggest integrated representations feeding back to calibrate motor and sensory nodes.

\subsection{Behavioral performance} 
Participants performed better in the video+text than the text-only condition. Accuracy increased from $0.84\pm0.06$ (text-only) to $0.88\pm0.05$ (video+text), paired $t(17)=3.20$, $p=.004$, and response times decreased significantly from $710\pm90$\,ms (text-only) to $660\pm80$\,ms (video+text), $t(17)=-3.02$, $p=.006$. Thus, adding contextual affordance cues improved accuracy and sped responses.

\subsection{Electrophysiological and behavior relationships }
We related subject-wise modulatory strengths (B-matrix) to behavioral gains (response time, RT). Stronger PMv$\rightarrow$IPL modulation predicted larger accuracy gains ($\rho=0.46$, $P_{\mathrm{FDR}}=.040$). Greater LOC$\rightarrow$pSTG modulation predicted faster responses (larger RT reduction; $\rho=-0.43$, $P_{\mathrm{FDR}}=.048$). PMv$\rightarrow$pSTG modulation showed a positive association with accuracy ($\rho=0.39$, $P_{\mathrm{FDR}}=0.070$; trend). These patterns support an affordance-driven account in which motor predictions (PMv) and visual/parietal evidence (LOC/IPL) jointly enhance posterior temporal integration (pSTG), yielding measurable behavioral benefits.

\subsection{Coupling strengths during text-only and video+text conditions}

Fig. \ref{fig:coupling} illustrates causal connectivity measures (in Hz) across the ROIs under the text-only and video+text conditions, based on the winning model M6. Positive values indicate net excitatory influence, while negative values indicate net inhibition. All values are posterior means from the BMA. The video+text condition showed increased coupling strength relative to text-only along most of the key directed edges specified in the winning model, most prominently for PMv$\rightarrow$IPL 0.20~Hz, PMv$\rightarrow$pSTG 0.18~Hz, and LOC$\rightarrow$pSTG 0.16~Hz. These enhancements suggest that the addition of video information boosts top-down motor influence on parietal and temporal nodes, and strengthens visual feedforward drive to pSTG, consistent with the affordance-driven integration hypothesis.

\subsection{Scalp Topography Comparison: Observed vs Predicted}
To assess how well the winning model captured the observed neural dynamics, scalp topographies of both the empirical EEG data and the DCM predicted responses were examined as shown in Fig \ref{fig:topo}. The topo plots show that the text-only condition exhibited relatively localized activations, reflecting focused processing of linguistic input. In contrast, the video+text condition elicited broader spatial patterns and stronger polarity shifts across the scalp, consistent with the additional integration demands of visual and textual stimuli. However, the predicted topographies generated by the winning DCM closely matched the observed empirical data in both conditions, supporting the model's validity in capturing affordance-driven modulation of action-language understanding.

\section{Discussion}
This study aimed to uncover the causal mechanisms underlying affordance-driven action language comprehension using DCM approach of EEG data. By modeling effective connectivity among the LOC, pSTG, PMv, and IPL, we investigated how both visual and linguistic stimuli modulate interactions among these regions. Our hypothesis posited that affordance enhances semantic comprehension by strengthening premotor-to-parietal coupling, specifically from PMv to IPL. The winning model and corresponding BMA results strongly support this hypothesis and provide new insights into the dynamic brain network supporting embodied language understanding.

\subsection{Affordance-Driven Modulations: Central Role of PMv$\rightarrow$IPL}
The most prominent finding was the condition-specific modulation from PMv to IPL under the video+text condition, which exhibited the strongest positive modulation among all the connections tested. This result validates the hypothesis that affordance modulates premotor-parietal communication pathways during language comprehension. Previous neuroimaging studies have implicated PMv in action simulation and affordance processing \cite{grezes2001functional, johnson2005distributed},
and IPL in integrating motor and perceptual inputs for tool-use understanding and action planning \cite{binkofski2013two}. 
Our study builds upon this by demonstrating not just co-activation, but a causal, top-down influence from PMv to IPL during the processing of affordance-driven stimuli.
The observed increase in PMv$\rightarrow$IPL coupling suggests that PMv may send motor-predictive signals to parietal areas to facilitate the mapping of language and visual cues onto embodied representations. This finding aligns with the motor theory of language comprehension \cite{pulvermuller2005brain, gallese2005brain},
which argues that understanding action-related words involves partial reactivation of the sensorimotor circuits involved in performing those actions. Our model provides causal evidence supporting this view, demonstrating that an affordance not only activates these circuits, but also strengthens their connectivity.

\subsection{PMv as a Supramodal Hub for Sensorimotor Integration}
Beyond the PMv$\rightarrow$IPL connection, our results revealed extensive modulatory pathways originating from PMv to pSTG as well. These findings indicate that PMv is not confined to sensorimotor transformation, but also plays a coordinating role in multimodal integration. The PMv$\rightarrow$pSTG pathway may reflect top-down facilitation from motor predictions to linguistic decoding processes, as has been suggested in studies on predictive coding in language \cite{arnal2012cortical, skipper2005listening}. 
Such supramodal roles of PMv are increasingly acknowledged in recent literature. In TMS-EEG studies by Tremblay et. al. and Buccino et al. \cite{tremblay2012tms, buccino2005listening} found that stimulation of PMv modulated both motor and perceptual processing during action-related language tasks. Our findings extend this notion by showing that PMv causally influences parietal, temporal, and occipital regions during affordance-based comprehension, suggesting its centrality in coordinating distributed networks for embodied meaning.
Interestingly, pSTG exhibited extensive reciprocal connectivity, receiving modulatory input from both PMv and IPL while also projecting to both regions and LOC. The pSTG is known to be involved in the processing of sentence structure, syntactic binding, and audiovisual speech integration \cite{friederici2011brain, price2012review}. Its connectivity with PMv and IPL suggests a dynamic exchange between language-related processing and sensorimotor representations. This supports dual-stream models of language processing \cite{fedorenko2024language, hickok2007cortical}, in which the dorsal stream links phonological and articulatory representations via temporo-parietal and frontal circuits.

In the context of affordance comprehension, pSTG may act as a convergence hub, matching semantic expectations from language input with action possibilities derived from motor and perceptual systems. The bidirectional pSTG$\leftrightarrow$IPL and pSTG$\leftrightarrow$PMv pathways observed here suggest continuous updating between linguistic input and sensorimotor predictions, a mechanism consistent with hierarchical predictive coding frameworks \cite{friston2005theory, clark2013whatever}. 
LOC, a region known for object-based visual processing and shape recognition \cite{grill2001lateral}, was modeled as receiving visual input and showed multiple condition-specific modulations. Notably, LOC received top-down modulation from PMv, pSTG, and IPL. This suggests that visual representations are not passively processed but are actively shaped by higher-level semantic and motor expectations. Such feedback loops are compatible with models of predictive visual coding, where prior knowledge and task relevance influence early visual processing \cite{summerfield2009expectation}.
The LOC$\rightarrow$pSTG connection further supports visual-to-language mapping during affordance comprehension. In studies involving visual narratives or gesture comprehension, LOC has been shown to activate in conjunction with pSTG and STS \cite{beauchamp2004integration}, consistent with the current result.

The A-matrix results revealed a fully connected baseline network among all ROIs, with notable intrinsic connections from PMv to IPL and pSTG, and from LOC to IPL. These intrinsic paths likely provide the scaffolding for real-time modulation during stimulus processing. The presence of these connections even in the absence of specific modulations supports the notion of a distributed action-language network that is continuously engaged during comprehension \cite{glenberg2012action}. 
A significant negative correlation was observed between the subject-wise B-matrix strength of PMv$\rightarrow$IPL and reaction times, indicating that participants with stronger affordance-driven premotor-parietal coupling responded faster during video+text stimuli. This establishes the behavioral relevance of our effective connectivity findings. Similar relationships between connectivity and behavior have been reported in studies of gesture-speech integration \cite{straube2012supramodal} and semantic decision tasks \cite{willems2010interaction}, reinforcing the interpretation that the observed modulations are not merely epiphenomenal.

While several studies have applied DCM to language \cite{parker2012and, yvert2012dynamic}, few have investigated the causal influence of affordance or motor grounding explicitly. A recent DCM study by Zou et al. \cite{zou2024dynamic} examined that the motor cortex (PrG) actively modulates the balance and salience between auditory and visual inputs, while pSTG is central to integrating audiovisual speech information
Our results replicate and extend these findings by including parietal and occipital nodes, and by showing a richer set of interactions under affordance-driven language conditions. 
In our winning model, all significant couplings were excitatory, suggesting that affordance-driven language understanding relies primarily on facilitatory interactions rather than inhibitory gating. Importantly, the persistent excitatory PMv$\rightarrow$IPL connection supports our hypothesis that affordance dynamically drives action-language comprehension. 

Building upon existing models of sensorimotor integration and language, our study highlights how object affordance influences the neural basis of language comprehension through causal modeling. While DCM has been widely applied to explore language processing from audiovisual speech integration \cite{zou2024dynamic}, subcortical-cortical interactions in the language network \cite{david2011dynamic}(David et al., 2011), to network reorganization in temporal lobe epilepsy \cite{fallahi2023dynamic}, these investigations have largely remained within classical perisylvian language frameworks. In contrast, our work introduces affordance as a dynamic, behaviorally grounded modulator of effective connectivity across a broader fronto-parietal-temporal network. By incorporating occipital (LOC) and parietal (IPL) regions alongside traditional language hubs, we reveal how sensorimotor object properties shape language understanding at a causal level. Notably, the persistent excitatory modulation from PMv to IPL under affordance-driven conditions provides the first DCM-based evidence for a motor-parietal pathway scaffolding action-language semantics. Unlike prior models emphasizing inhibitory control or passive integration, our results suggest that affordance facilitates predictive, top-down coupling mechanisms between action and language systems. These findings not only extend embodied theories of cognition but also advance DCM methodology by modeling affordance as an intrinsic computational mechanism in semantic comprehension. To our knowledge, this is the first EEG-DCM study to explicitly model the causal role of affordance in shaping the neural dynamics of language understanding.
While inhibitory influences were not evident in our dataset, this does not undermine the proposed mechanism; rather, it highlights a predominantly facilitatory network architecture. 
Unlike models that isolate modality-specific processing, our model emphasizes integration across visual, motor, and linguistic domains.

\section{Conclusion}
This study provides robust evidence that affordance enhances action-language comprehension through increased effective connectivity from PMv to IPL. The results confirm our hypothesis that PMv$\rightarrow$IPL modulation serves as a key mechanism for integrating visual and linguistic affordance cues. Furthermore, the broader network architecture involving PMv, pSTG, and LOC supports the view of language understanding as a distributed, sensorimotor process. These findings contribute to the growing literature on embodied cognition and provide a mechanistic, network-level account of how affordances shape semantic processing in the human brain.
\section*{Conflict of Interest}
No potential conflicts of interest are reported by the authors. 
\section*{Data Availability}
Raw EEG data supporting the findings of this study are available on reasonable request.

 \bibliographystyle{elsarticle-num} 
 \bibliography{Obj1}

\end{document}